# Walking Through an Exploded Star:
# Rendering Supernova Remnant Cassiopeia A into Virtual Reality

Arcand, K.K.[1], Jiang, E.[2], Price, S.[1], Watzke, M.[1], Sgouros, T.[2], Edmonds, P.[1]
([1]Smithsonian Astrophysical Observatory/Chandra X-ray Observatory, [2]Brown University)

**ABSTRACT:**

NASA and other astrophysical data of the Cassiopeia A supernova remnant have been rendered into a three-dimensional virtual reality (VR) and augmented reality (AR) program, the first of its kind. This data-driven experience of a supernova remnant allows viewers to "walk" inside the leftovers from the explosion of a massive star, select the parts of the supernova remnant to engage with, and access descriptive texts on what the materials are. The basis of this program is a unique 3D model of the 340-year old remains of a stellar explosion, made by combining data from NASA's Chandra X-ray Observatory, Spitzer Space Telescope, and ground-based facilities. A collaboration between the Smithsonian Astrophysical Observatory and Brown University allowed the 3D astronomical data collected on Cassiopeia A to be featured in the VR/AR program - an innovation in digital technologies with public, education, and research-based impacts.

## 1. INTRODUCTION

### 1.1. Overview of Virtual Reality (VR)

Virtual Reality (VR) is computer technology that simulates a user's physical presence in a virtual environment. (VR's close relative, Augmented Reality (AR), adds elements, such as text, overlays and audio, to enhance that experience with sensory input and is briefly discussed in section 1.3). VR has been around as an idea since the 1960's, and a reality in some form since the 1980's (Faisal, 2017). Though it has faced many starts and stops (Stein, 2015), e.g., the promise of VRML (Virtual Reality Markup Language) in the late 1990's[1], VR has become more commonplace in the consumer market since about 2010 (Faisal, 2017). With the promise of improving the gaming industry, media, and even adult entertainment, there are major commercial driving forces behind the technology's development (*Virtual Reality Payments* 2016; *Virtual Reality 101* 2016).

The increased commercial prominence of these technologies, including the availability of less expensive but still good quality and more user-friendly technologies, provides opportunity for science as well (Ferrand et al., 2016). In science, there is a potential for VR to revolutionize how experts visualize and analyze their data, and also how that data is then communicated to non-experts, from molecular modeling to environmental conservation (Isenberg, 2013).

In medicine alone, VR offers a unique tool for data visualization and comprehension as well as continuing education and user experiences. Programs in progress range from improving health workers' understanding of brain damage (Hung et al., 2014), to implementing virtual surgery training for medical students (Murphy, 2018), to applying VR and AR techniques in an accessible way to the treatment of Alzheimer's disease (Garcia-Betances et al., 2015). Virtual reality has been shown to improve upon the traditional tangible model of using dummies to enhance medical students' preparation for assisting patients in higher risk, real life scenarios. It might allow physicians and other health professionals to work with individually developed cases to practice applicable caregiving skills, such as looking at and speaking to new patients (Murphy, 2018).

Beyond the universe of the body, and out to the Universe at large, astronomical data sets often offer high-resolution, multi-wavelength, multi-dimensional (lately, even multi-messenger) information. The process of converting photons, or packets of energy, into 2D images has been documented and studied (Rector et al., 2017; Arcand et al., 2013; DePasquale et al., 2015; Rector et al., 2007) but the translation of that information into 3D forms that takes advantage of human perspective, cognition and stereoscopic vision, less so (Ferrand et al., 2016). Since the Universe is multidimensional itself, as Fluke and Barnes (2016) ask: 'Are we making the best use of the astronomer's [and non-expert's] personal visual processing system to discover knowledge?'

A selection of astronomical VR experiences the authors have come into contact with range from exploring exoplanets (planets outside our solar system) through science-informed 3D artists' impressions converted into VR[2], to experiencing NASA mission spacecraft in VR such as the James Webb Space Telescope[3], to walking across the surface of Mars[4]. The authors' experience also includes exploring dozens of massive stars from the perspective of the supermassive black hole at the center of our Galaxy[5] (Russell, 2018) and viewing radio data cubes of a spiral galaxy (Ferrand et al., 2016).

**1.2. Data Path for Cassiopeia A: From 2D to VR**

Supernova explosions are among the most violent events in the universe. When the nuclear power source at the center of a massive star is exhausted, the core collapses and in less than a second, a neutron star typically forms. This process releases an enormous amount of energy, which reverses the implosion, blows material outwards and produces a brilliant visual outburst. The resulting debris field is referred to as a "supernova remnant."

Cassiopeia A (Cas A) is a supernova remnant from an explosion that occurred approximately 340 years ago in Earth's timeframe (see Figure 1). A multi-wavelength three-dimensional (3D) reconstruction of this remnant was created using X-ray data from *Chandra*, infrared data from *Spitzer* and optical data from ground-based telescopes. In Figure 1, the green regions are mostly iron, the yellow regions include a combination of argon and silicon, the red regions are cooler explosion debris, and the blue regions show the outer blast wave.

When elements created inside a supernova are heated they emit light at specific wavelengths. Because of the Doppler effect, elements moving towards the observer will have shorter wavelengths and elements moving away will produce longer wavelengths. Since the amount of the wavelength shift is related to the speed of motion, the velocity of the debris can be determined. By combining this Doppler information with the expectation that the stellar debris are expanding radially outwards from the explosion center, Delaney et al. (2010) applied simple geometry to construct a 3D model of Cas A (Figure 2). A program called 3D Slicer - modified for astronomical use by the Astronomical Medicine Project at Harvard[6] - was used to display and manipulate the 3D model.

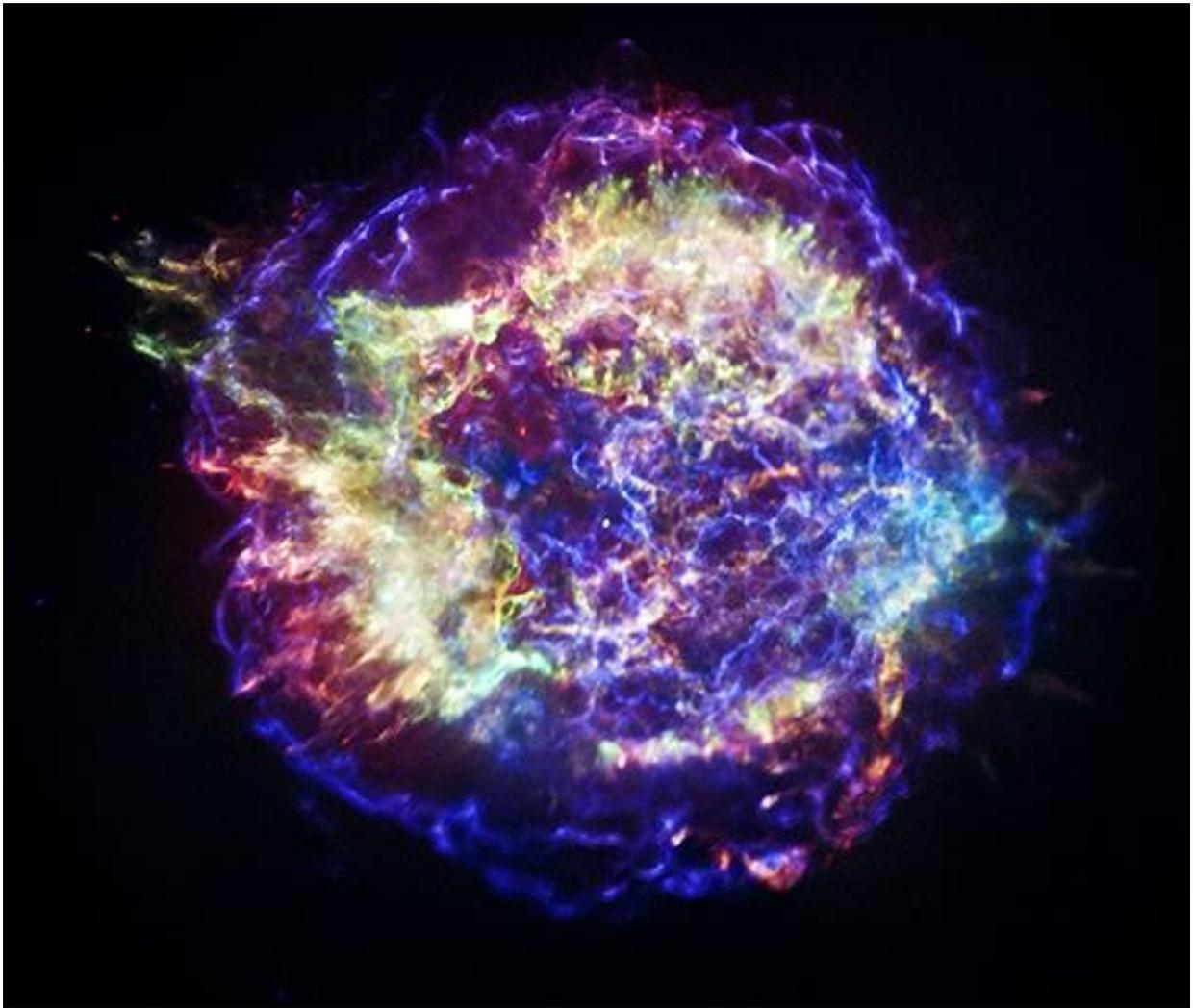

Figure 1. Cassiopeia A (Cas A) is a supernova remnant located about 10,000 light years from Earth. This 2D visual representation of Cas A has been processed to show with clarity the appearance of Cas A in different bands of X-rays. This will aid astronomers in their efforts to reconstruct details of the supernova process such as the size of the star, its chemical makeup, and the explosion mechanism. The color scheme used in this image is the following: low-energy X-rays are red, medium-energy ones are green, and the highest-energy X-rays detected by Chandra are colored blue. The image is 8.91 arcmin across (or about 29 light years). Credit: NASA/CXC/SAO

This visualization[7] shows that there are two main components to Cas A: a spherical component in the outer parts of the remnant containing light elements like helium and carbon from the outer layer of the exploded star, and a flattened (disk-like) component in the inner region containing heavier elements like argon and iron from the inner layers of the star. The blue filaments defining the blast wave show a different type of radiation that does not emit light at discrete wavelengths, and, therefore, were not included in the 3D model.

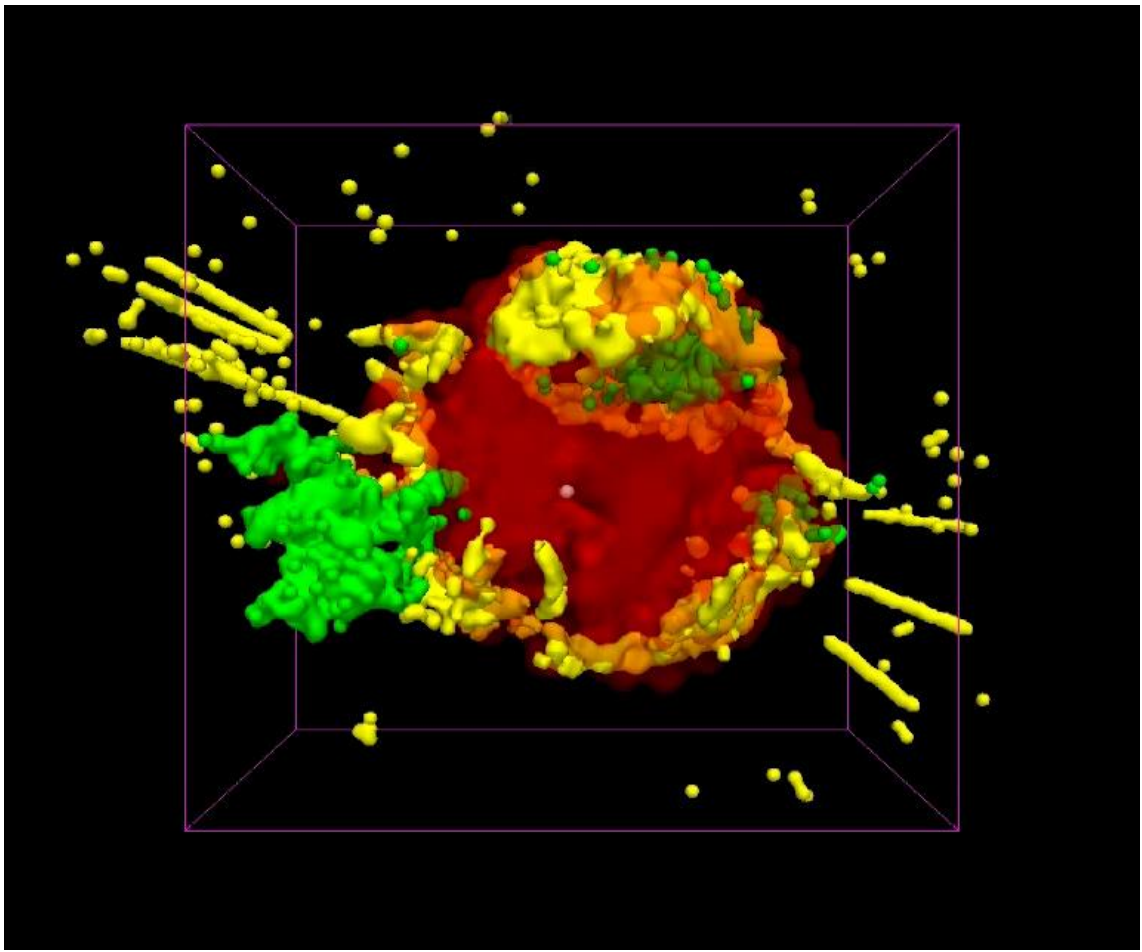

**Figure 2. By combining data from Chandra, the Spitzer Space Telescope, and ground-based optical observations, astronomers were able to construct the first three-dimensional fly-through of a supernova remnant. This 3D visualization (shown here as a still image) was made possible by importing the data into a medical imaging program that has been adapted for astronomical use. Commercial software was then used to create the version of the 3-D data. The green region shown in the image is mostly iron observed in X-rays; the yellow region is mostly argon and silicon seen in X-rays, optical and infrared and the red region is cooler debris seen in the infrared. The positions of these points in three-dimensional space were found by using the Doppler effect and simple assumptions about the supernova explosion. Credit: NASA/CXC/MIT/T.Delaney et al.**

High-velocity jets of this material are shooting out from the explosion in the plane of the disk-like component mentioned above. Jets of silicon appear in the upper left and lower right regions, while plumes of iron are seen in the lower left and northern regions. These had been studied before the 3D model was made, but their orientation and position with respect to the rest of the debris field had not been mapped before this model.

The insight into the structure of Cas A gained from this 3D visualization is important for astronomers who build models of supernova explosions. They have learned that the outer layers of the star come off spherically, but the inner layers come out more disk-like with high-velocity jets in multiple directions (Delaney et al., 2010). Since the Delaney et al. (2010) study, two other groups have constructed 3D models of Cas A (Milisavljevic and Fesen, 2015; Orlando et al., 2016), demonstrating the rich scientific value of such visualizations for astronomers.

This 3D visualization was initially created as an interactive for desktop viewing and interaction[8]. To move beyond the small screen, it was next translated into a 3D printable format[9] (Arcand et al., 2017), a format which is particularly conducive for exploration by blind and visually impaired populations (Grice et al., 2015; Christian et al., 2015).

The authors considered Virtual Reality (Figure 3) as the next step in the translation process, an additional experience beyond the 2D visual and 3D tactile ones (Eriksson, 2014).

**1.3. Augmented Reality in Astronomy**

As mentioned above, Augmented Reality (AR) adds text, image, sound-based elements or other effects to deliver an enhanced view for the user experience with additional sensory input, typically in a merging of real or "live" and virtual information. AR has been rising in popularity, with the number of smartphone, wearable computing devices, and other similar smart devices being pushed in the consumer

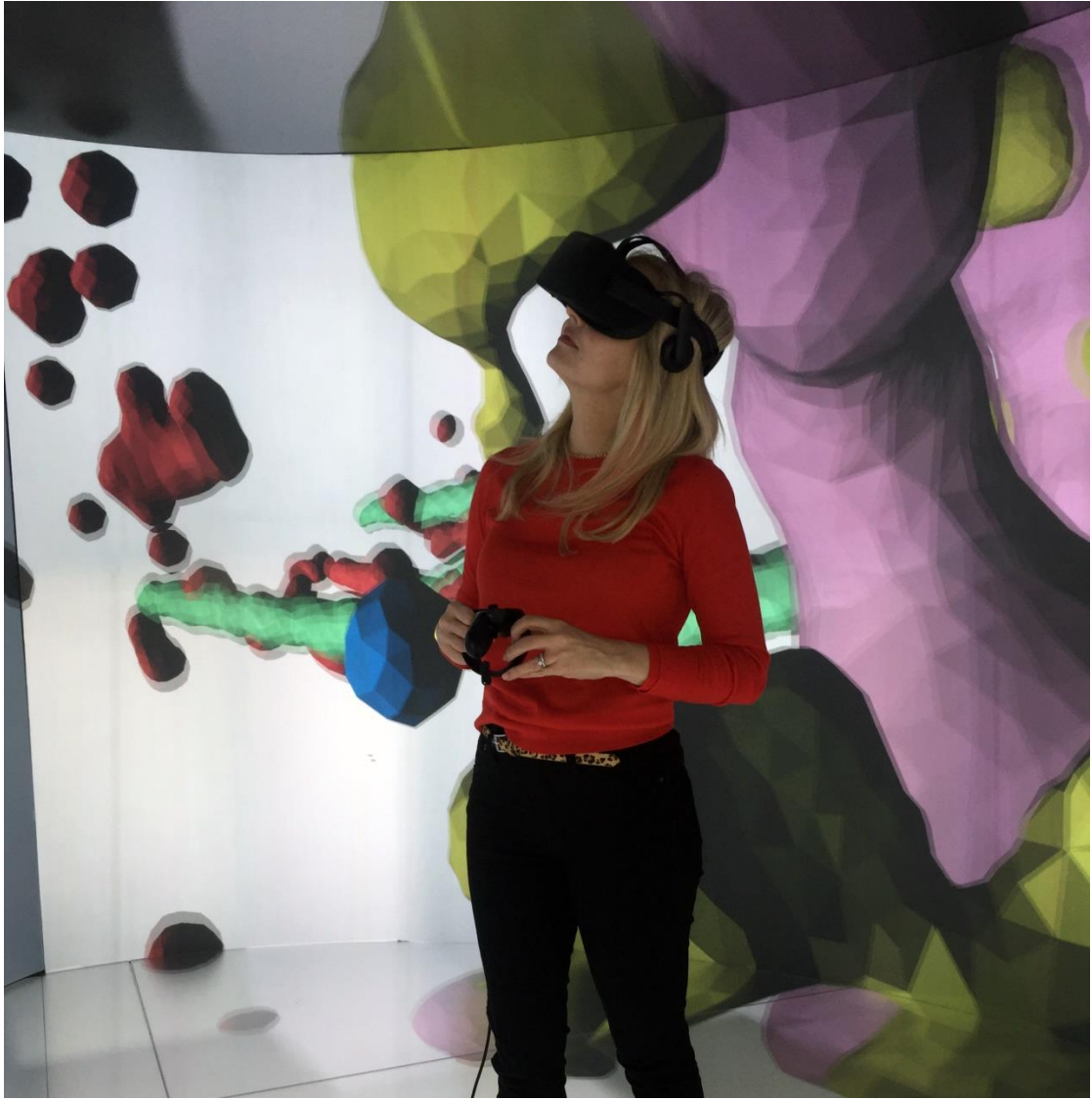

**Figure 3. A three-dimensional virtual reality (VR) with augmented reality (AR) version of the 3D data of Cas A allows the user to walk inside the debris from a massive stellar explosion, select the parts of the supernova remnant to engage with, and access short captions on what the materials are. This photo was taken of the first author inside the Brown University YURT, or VR CAVE, during testing of the Oculus Rift hardware and application. Credit: NASA/CXC/SAO/E.Jiang**

market (Vogt & Shingles, 2013[10]). From the "computational interfaces" popularized by Pokemon Go[11] (Sicart, 2017), to AR leopards that visitors can interact with to help promote conservation themes[12], AR is expected to continue rising in popularity in the next few years[13].

AR is, therefore, neither a completely new nor a completely unexpected area for topics in astronomy. Vogt and Shingles (2013) for example[14], make note of AR-based designs in astronomy for education purposes including 3D displays of our Solar system, 3D demonstrations of the Sun-Earth interaction, and a NASA application that allows users to get 'up close and personal[15]' with NASA spacecraft merged with AR overlays. Vogt and Shingles also include the use of AR for astronomy research, citing their own interactive AR figure of the supernova remnant N132d.

## 2. VR TRANSLATION/TECHNICAL SPECIFICATIONS

Currently, virtual reality software has not been standardized to the point where we can use data as direct input into a VR program. While software exists that allows visualizations to be displayed across different platforms (caves[16], PowerWalls[17], 3DTVs[18], head-mounted displays[19]) using a variety of input devices (6 degree-of-freedom trackers[20], multi-touch input devices[21], haptic devices[22]), we have yet to design a system that can support raw data without any external software. Therefore, with each new type of dataset, effort must be made to build the bridge between raw data and VR software in order to produce the visualization.

Given volumetric data (where volume, or the data inside an object, is rendered - as in an MRI or CT scan in medicine) and polygonal data (where only the surface, or the outside data, is rendered) on the supernova remnant Cassiopeia A collected by *Chandra*, *Spitzer*, and ground-based optical observatories, this formatting of Cas A employs volume and surface rendering techniques using the Visualization Toolkit (VTK[23]) to create a MinVR[24]-enabled program.

### 2.1. METHODS

#### 2.1.1. Volume and Surface Rendering

To render the supernova both volumetrically and polygonally, we used VTK, an open-source, freely available software system for 3D computer graphics, image processing, and visualization. VTK supports a wide variety of visualization algorithms including scalar, vector, tensor, texture, and volumetric methods, as well as advanced modeling techniques such as implicit modeling, polygon reduction, mesh smoothing, and contouring (https://www.vtk.org; Avila, Kitware, 2010). VTK has a suite of 3D interaction widgets.

With VTK, we were able to read in the supernova datasets and use its built-in filters and mappers to render the remnant's volume and surface. The remnant is composed of seven different parts, and each part is represented by a different color in the surface model.

### 2.1.2. Integrating with MinVR

MinVR is an open-source project developed and maintained collectively by the University of Minnesota, Brown University, and Malcalester College. It aims to support data visualization and virtual reality research projects by providing a cross-platform VR toolkit that can be used in many different VR displays, including Brown University's YURT (Yurt Ultimate Reality Theatre, or CAVE).

One technical challenge of this project was integrating the VTK program with MinVR. Since both programs had their own render function, we had to use VTK's external module to allow the VTK program to accept an external render window and render loop. As has been the case when working with our 3D/VR projects thus far, multiple bridges need to be built between technologies in order to create something new. We worked with the VTK to improve their external camera capacity. In the end, we were able to complete the integration and display the supernova models in the YURT.

### 2.2. PROGRESS & RESULTS - VOLUME AND SURFACE RENDERING

**Figure 4. Volume Rendering**

(a)

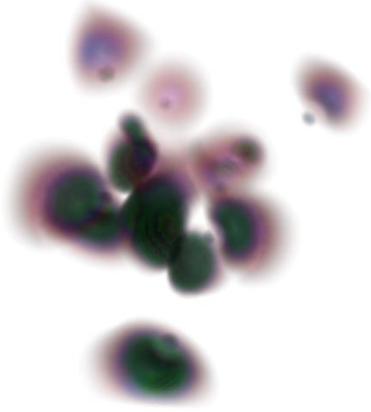

(b)

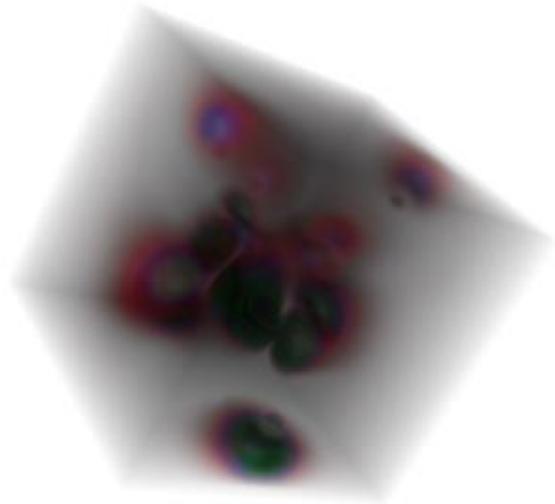

(c)

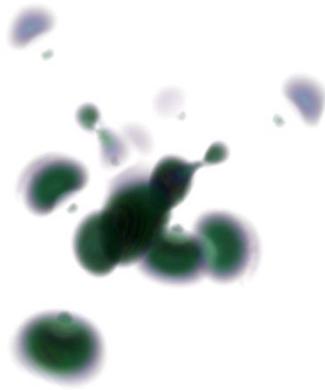

Figure 4(a) shows a sample volume rendering demo that uses iron density data to simulate what a supernova would look like given its volumetric data.

Figure 4(b) illustrates that when the opacity level is low enough, one can observe an outer cube that encompasses the iron density data, illustrating the importance of adjusting opacity levels when volume rendering.

On the other hand, when the opacity level is too high, we cannot see enough of the data to observe the differing measurements of density. Figure 4(c) shows that we must work to find an optimal opacity level and an intuitive color scale.

**Figure 5. Surface Rendering**

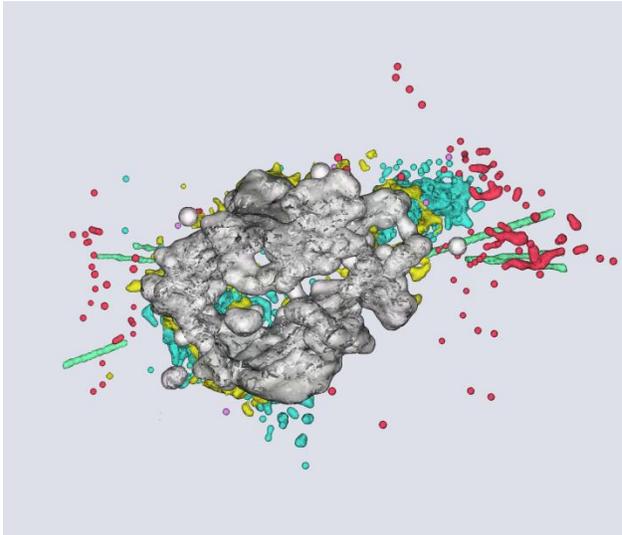
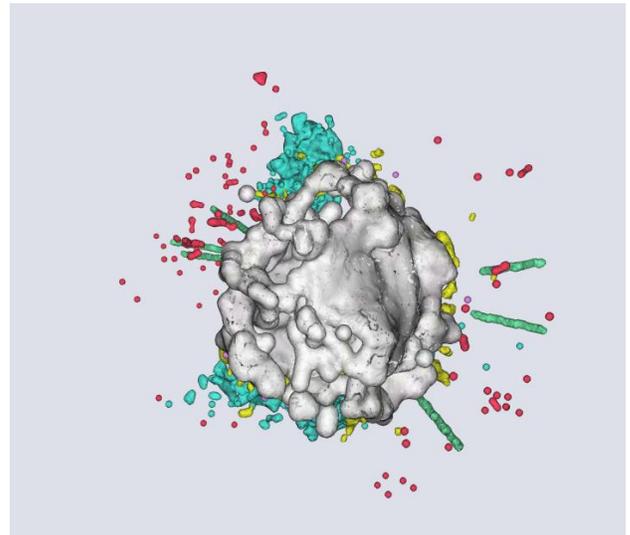
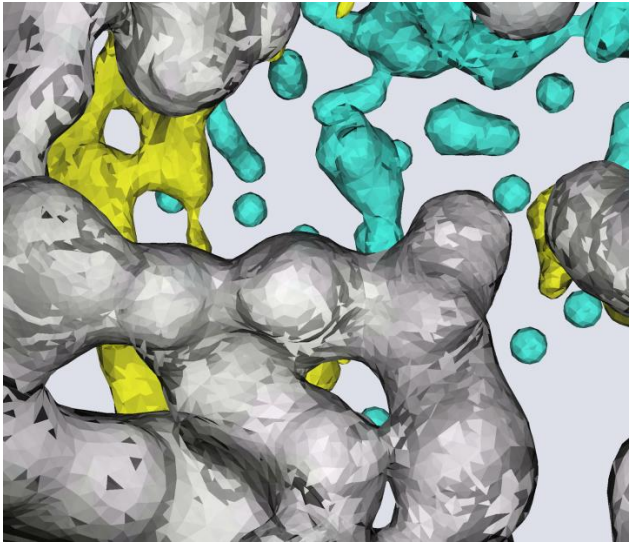

The model in Figure 5(a) is a surface rendering of Cas A, which is made up of 7 different parts, as shown in different colors. Each part is a separate ASCII data file consisting of polygon data and triangular strips.

The 7 parts shown in Figure 5(b) include a spherical component (purple), a tilted thick disk (gray), and multiple ejecta jets/pistons (green) and optical fast-moving knots (red, yellow, blue, pink) all populating the thick disk plane.

Figure 5(c) shows the view from inside the spherical structure of the supernova. Here, it is possible to see that the rendering is a mesh of triangles that shape the surface.

### 2.3. Augmented Reality: Narrative Additions

For the Cas A VR experience, adding interactive text over the VR object was an important narrative component of the overall experience due to the complexity of the science model. Providing contextual information has been shown to improve the user experience in understanding and enjoyment of 2D and 3D astronomy images for both experts and non-experts, and across technological platforms (Smith et al., 2017, 2017, 2014 & 2010). Therefore expanding the addition of contextual information to VR data sets seemed ideal for users of all kinds.

Our enhanced user experience of the VR Cas A model includes annotations for each of the parts of the supernova remnant (noting, for example, the neutron star, the iron and silicon debris, etc.) to describe both its components and its overall structure. Users can select a specific part of the supernova remnant by using their input device or wand (such as the Oculus Touch) to access the annotations and bring them up in their VR environment. They can cycle through each notation to discover more information about Cas A. These additional interactive narrative features may also help educators to more effectively tell the life story of a star, and provide resources for researchers who are observing the changes in size, density, and shape of stars.

**Figure 6. Augmented Reality**

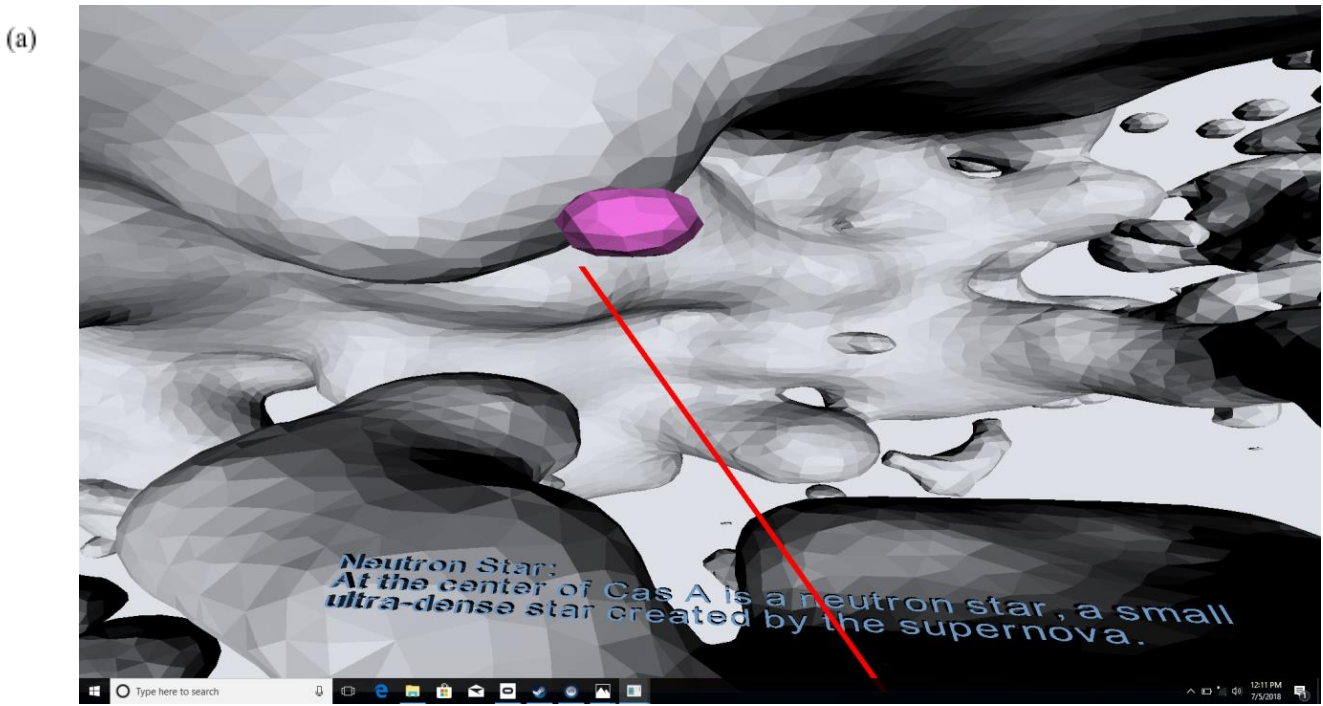

Figure 6(a) shows the addition of narrative text, where the user can select a part of Cas A to focus on and access captions. In this case, the user has chosen to focus on the Neutron Star at the center of the remnant.

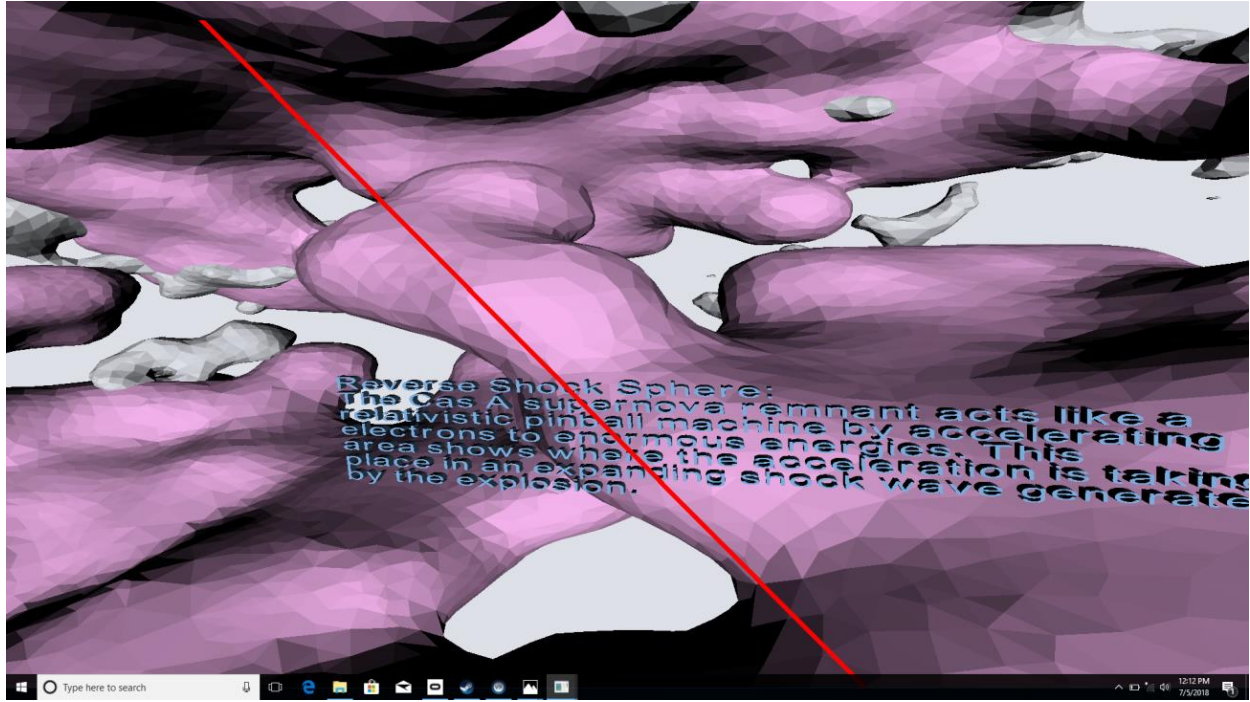

Figure 6(b) captures the screen of a user highlighting the Reverse Shock Sphere that demonstrates wave expansion. Note how the caption superimposed over the image uses analogy to help potentially increase understanding (Smith, et al 2017).

## 3. DISCUSSION

The VR Cas A experience was created for 3D immersive environments such as CAVES and the Oculus Rift (Clark, 2014). Adaptations for cell phone usage with pop up personal VR viewers such as Google Cardboard, and that can also be used in a browser without any VR viewers, have been created to allow additional entry points for viewers who do not have access to more expensive equipment (for an online demonstration see http://chandra.si.edu/vr/casa).

The Cas A 3D VR model has the potential to be a useful tool in engaging experts and non-experts in the data of astronomy and applications of computer science. For non-experts, specifically, astronomy data in general are popular as a science topic, as evidenced by the ubiquitous placement of astronomical images throughout popular culture everywhere from bed linens to computer wallpaper (and wallpaper murals).[25]

By linking the data and images of Cas A with unique computer tools in a project such as this, new connections can be made. Astronomy models in virtual reality can provide an unexpected visual and perceptual palette for the modern viewer. Such applications may be able to assist participants in establishing a sense of presence with data that is, due to the nature of the distance from Earth, sheer scale, and other factors, otherwise difficult to relate to. This data can be shown at a monumental scale with VR. There are difficulties with VR technologies to consider, however, such as visual disconnect causing motion sickness, the need to incorporate an illusion of boundless movement, accessibility (both for underserved socioeconomic areas and also for physical accessibility by people who are visually impaired)[26], as well as struggles in establishing touch-responsive features (Steinicke, 2016; Amer & Peralez, 2014).

Just as in video games and educational software, VR-enabled science requires narrative integration whether for expert or non-expert users.[27] The key to making such a project more than just a toy or gee-whiz moment is to provide meaningful information and content that is clearly embedded in the virtual experience. Projects such as this Cas A 3D VR/AR model could be used as a "jumping off" point for opportunities in a host of topics in astrophysics, chemistry, computer science and more.

With current market trends for equipment and more adaptations of content and technologies for VR experiences (including multimodal access points for those with physical disabilities), it is a critical time to create quality astronomy-based materials for experienced VR users, as well as those new to VR. In education specifically, helping students maximize the potential of VR could be done partially through output adaptations to platforms such as YouTube which can host 360 degree versions of many VR videos where they can act as a canvas for individualized VR experiences tailored to the needs of different learners (Cotabish, 2017). This aspect of the viewer-driven experience in VR broadly (Chen et al., 2014) could potentially positively impact users with different learning styles, viewers with autism (Lahiri et al., 2015), participants with different physical abilities or other special needs (Tyler-Wood et al., 2015), and more.

**3.1. NEXT STEPS**

**3.1.1. Expansion of Astrophysics VR Library**

While our current results have provided an immersive and interactive rendering of the supernova remnant Cassiopeia A, we plan to expand our astrophysics VR data sets to build additional 3D visualizations to help illustrate more fully the life-cycle of stars, from birth to death, with further interactivity included for the user. We are investigating the application of additional 3D data-driven models to import into the VR pipeline described in previous sections. Future models include volumetric data files of supernovas and younger star systems that might also lead to more comprehensive models. We are also working to make the existing VR application ADA/Section 508 compliant for accessibility.[28]

**3.1.2. Beyond Supernovas & Supernova Remnants**

Through rendering 3D models of Cas A, this project has in addition implemented a generic program with examples of how to create similar programs to read in and display such data. In the future, the authors hope to demonstrate new models of another famous supernova, SN 1987A, an explosion on the surface of a white dwarf, V745 Sco, and other astrophysical density data. The goal is to generate these models with less effort than for Cas A. Our intent is that this generic program could act as a skeleton and tutorial for future data sets in biomedical, physical or other fields.[29]

**4. CONCLUSION**

We are excited about both the current abilities and the future potential of opportunities for using VR/AR in astronomy. There is a potential for unique educational experiences by wedding the popularity of a visual science like astronomy with the technological advances that continue to evolve in VR/AR. Making "real world" examples like Cas A part of the cadre of VR/AR could spur creative ideas of how to infuse science into a realm that might typically include more content from science fiction. We look forward to making deeper connections with the subject matter we are most familiar with, to see how we can expand its content into the virtual third dimension and beyond.

## ACKNOWLEDGMENTS

Many special thanks to the Virtual Reality Lab at the Center for Computation and Visualization at Brown University without which this project could not have been done. The Cassiopeia A digital 3D model was originally developed with Dr. Tracey Delaney of West Virginia Wesleyan College (formerly of MIT), with the Chandra X-ray Center at the Smithsonian Astrophysical Observatory, in Cambridge, MA, with funding by NASA under contract NAS8-03060.

---

[1] E.g., http://www.cnn.com/TECH/9710/14/3.d.reality.lat/index.html
[2] https://www.space.com/38749-visit-six-real-exoplanets-with-virtual-reality.html
[3] https://connect.unity.com/p/vr-experience-james-webb-space-telescope
[4] https://www.jpl.nasa.gov/news/news.php?feature=6978
[5] http://chandra.si.edu/photo/2018/gcenter360/
[6] Note: since archived, Initiative in Innovative Computing: http://am.iic.harvard.edu
[7] Cas A 3D Model video at: http://chandra.si.edu/photo/2009/casa2/

[8] Smithsonian 3D model: http://3d.si.edu/explorer?mid=45
[9] Printable Cas A 3D Model: http://chandra.si.edu/3dprint
[10] http://ieeexplore.ieee.org/document/6970345/?reload=true
[11] Sicart (2017) notes the potential drawbacks for AR, including the ability for it to be used by private companies and other for-profit actors that are not accountable to the physical spaces that they create virtual revenue within, which can reduce equity rather than provide a new platform for it.
[12] See for example https://www.smithsonianmag.com/travel/how-augmented-reality-helping-raise-awareness-about-one-armenias-most-endangered-species-1-180967670/
[13] See for example https://www.thenational.ae/business/peter-nowak-why-augmented-reality-will-be-a-big-trend-in-2017-1.32774; https://venturebeat.com/2018/02/08/the-nyt-is-boarding-the-ar-train-heres-what-that-means-for-storytelling/
[14] The Vogt and Shingles (2013) paper can be downloaded as an augmented article, illustrating the utility of the technology promoted in the paper. The authors argue for collaboration between science researchers and publishing platforms to create more stable, as well as backwards compatible, content for AR.
[15] Examples at https://www.nasa.gov/mission_pages/msl/news/app20120711.html
[16] Brown Center for Computation and Visualization CAVE: https://web1.ccv.brown.edu/viz-cave ;Juarez, A., Schonenberg, B., & Bartneck, C. (2010). Implementing a Low-Cost CAVE System Using the CryEngine2. *Entertainment Computing*, 1(3-4), 157-164. http://www.bartneck.de/publications/2010/caveCryEngine/ [Accessed 17 Mar. 2018]
[17] See for example the Viscon Virtual Reality VR PowerWall: http://viscon.de/en/vr-2/vr-powerwall/
[18] LG provides a striking example of a 3DTV: https://www.lg.com/us/tvs/lg-OLED55E6P-oled-4k-tv; Reasonably priced glasses for the 3DTV are available at Kmart: https://www.kmart.com/edimensional-ed-4-pack-cinema-3d-glasses-for-p-SPM8624650802?plpSellerId=Ami%20Ventures%20Inc&prdNo=2&blockNo=2&blockType=G2
[19] A very affordable version of the head mounted display is available at: Amazon.com. (2018). [online] Available at: https://www.amazon.com/AUPALLA-Cardboard-Christmas-Version-Virtual/dp/B01LZA1EKZ/ref=sr_1_15/143-5425711-5657840?ie=UTF8&qid=1521395844&sr=8-15&keywords=google%27s+cardboard [Accessed 5 Jul. 2018]; See for a more costly option: Amazon.com. (2018). [online] Available at: https://www.amazon.com/Oculus-Rift-Virtual-Reality-Headset-pc/dp/B00VF0IXEY/ref=sr_1_4?s=electronics&ie =UTF8&qid=1521395894&sr=1-4&keywords=oculus+rift [Accessed 5 Jul. 2018].
[20] https://www.roadtovr.com/introduction-positional-tracking-degrees-freedom-dof/
[21] Marzo, A., Bossavit, B., & Hachet, M. (2014). Combining Multi-Touch Input and Design Movement for 3D Manipulations in Mobile Augmented Reality Environments, *ACM Symposium on Spatial User* Interaction [short paper], Oct 2014, Honolulu, HI, United States. Available at: https://hal.inria.fr/hal-01056226/document [Accessed 13 Jul. 2018].
[22] See this video for an explanation of haptic devices: https://www.youtube.com/watch?v=ABeAAHF6k1k
[23] "Enabled Applications." *VTK*. Kitware, n.d. Web. http://www.vtk.org/
[24] "MinVR/MinVR." *GitHub*. Brown University, 20 Apr. 2017. Web. Retrieved from https://github.com/MinVR/MinVR
[25] See, for example, the Space and Astronomy section of *Geek Wrapped* for an assortment of products decorated with astronomical images: *Best Space and Astronomy Gifts* (2018), viewed 8 March 2018 <https://www.geekwrapped.com/astronomy>.
[26] The ADA National Network explains *What is the Americans with Disabilities Act?*, viewed 8 March 2018. <https://adata.org/learn-about-ada>.
[27] The necessity of narrative is discussed in Gottschalk, M. (2016) *Virtual Reality is the Most Powerful Medium of Our Time*.
[28] United States Access Board (2000) sets out the *Section 508 Standards for Electronic and Information Technology*.
[29] The CXC does not endorse any specific commercial product, including the virtual reality technologies referenced throughout this paper.